\documentclass[english,prb,preprint]{revtex4}
\usepackage[]{fontenc}
\usepackage[latin1]{inputenc}
\usepackage{graphicx}

\makeatletter

\usepackage{bm}

\usepackage{babel}
\makeatother
\begin{document}

\title{Thermal instability of an expanding dusty plasma with equilibrium
cooling}

\author{Madhurjya P. Bora and Manasi Buzar Baruah}

\address{Physics Department, Gauhati University, Guwahati 781014, India.}

\begin{abstract}
We present an analysis of radiation induced instabilities in an expanding
plasma with considerable presence of dust particles and equilibrium
cooling. We have shown that the equilibrium expansion and cooling
destabilize the radiation condensation modes and the presence of dust
particles enhances this effect. We have examined our results in the
context of ionized, dusty-plasma environments such as those found in
planetary nebulae
(PNe). We show that due to the non-static equilibrium and finite equilibrium
cooling, small-scale localized structures formed out of thermal instability,
become transient, which agrees with the observational results. The
dust-charge fluctuation is found to heavily suppress these instabilities,
though in view of non-availability of convincing experimental data,
a definitive conclusion could not be made.
\end{abstract}
\maketitle

\section{Introduction}

Thermal instability is long thought to be a relevant astrophysical
process which is responsible for existence of smaller, non-gravitational
condensations such as those found in solar prominences, interstellar
clouds, and planetary nebulae (PNe). Earlier works on thermal instability
include those of Parker\cite{pa-1}, Weymann\cite{we-1}, and Field\cite{fi-1},
among which, Field's seminal work on thermal instability is considered
to be a comprehensive treatise on the subject till date. Recently,
Gomez-Pelaez and Moreno-Insertis\cite{go-1} have addressed the process
of thermal instability in a gravitating medium undergoing expansion.
Several other authors have considered occurrence of thermal instability
in different contexts and parameter regimes. An analysis of the thermal
instability of an optically thin plasma in the nonlinear regime is
considered by Steele et. al. \cite{ste-1} Some of the very recent
works involving numerical simulations, address the issues of dense
structure formation in a hot protogalactic environment \cite{baek-1,baek-2}
and the interplay between thermal and magnetorotational instability
(MRI) in interstellar media \cite{pointek-1,pointek-2}.

It is also now well known that the dust particles constitute an ubiquitous
and important component of many astrophysical plasmas including interstellar
clouds, stellar and planetary atmospheres, planetary nebulae, giant
H-II regions (GHR), hot interstellar matter (ISM), etc. The presence
of dust in an ionized astrophysical structure, besides being responsible
for introduction of new wave modes related to the dust dynamics, can
also significantly modify the thermal structure of the plasma \cite{sh-1,do-1}.
One of the important ways dust can modify the stability properties
is by capturing energetic electrons and cooling through radiation\cite{ol-1,ba-1}
in the infrared region and we expect that the presence of dust particles
modify the corresponding growth rates significantly. The charge fluctuation
dynamics of the dust particles is also found to introduce new acoustic
modes in the plasma \cite{jana,bora}. Besides, charging of the dust
grains makes them an integral part of the multi-component plasma which
also helps maintain the colloidal nature of the plasma. Apart from
the physical influence of the presence of dust particles in such astrophysical
environments, the scenario becomes more complicated when one considers
the dynamic behavior of the background. For example, the planetary
nebulae are known to have severe equilibrium expansions, in many cases
with supersonic expansion velocities \cite{phillips-1,schon-1}. It
is therefore of particular interest to review the process of condensations
in ionized plasmas taking into consideration the effect of the dust
dynamics and their participation in the overall stability with a non-static
background. In this work, we have undertaken a systematic study of
the thermal condensation process in such an ionized plasma environment
which undergoes equilibrium expansion and cooling having a considerable
presence of dust particles which constitute another charged species
of the plasma. We would like to view the outcome of this analyses
in the context of condensations observed in planetary nebulae (PNe).

The problem of existence of dust particles inside the hot ionized
environment of a PNe is yet to be understood fully \cite{sta-1}.
It has however been proposed that majority of the dust grains could
be destroyed in the hot ionized regions of plasma in a PNe, as it
evolves, either because of spallation by hard UV photons or shocks.
Subsequently, a PNe or a GHR should be free of dust grains \cite{pot-1,lenz-1}.
However, in many cases the destruction of dust particles inside an
ionized environment is not complete. For example, recent observations
of the infrared spectra of the evolved nebula NGC 6445\cite{ki-1,ki-2},
it has been shown that only a little destruction of the dust grains
inside the ionized region of the nebula could have occurred and dust-grain
separation inside the ionized region is not plausible\cite{va-1}.
Even earlier attempts to detect Ca lines in PNes suggest that Ca is
probably locked in the dust grains present inside the ionized environment
\cite{volk-1,fi-2}. Theoretical investigations, so far, indicate
that dust particles are gradually accelerated to higher and higher
velocities by the radiation pressure and are being expelled from a
nebula \cite{okro-1,mar-1}, which however do not take into account
the mechanism of charging of the dust grains. The charging of the
dust grains freezes them into the ionized gas \cite{sta-1}, about
which we have already referred. 

In Section 2, we describe our model equations with background cooling
and expansion. In Section 3, we apply the linear perturbation theory
to examine the thermal condensation modes. In Section 4, we consider
the case of static dust charge with the help of a WKB formalism and
derive the linear dispersion relation. We numerically examine the
dust-charge fluctuation dynamics in Section 5. Finally, we summarise
our observations in Section 6.

\section{Model equations}

Below, we write down the equations for a weakly collisional, fully
ionized dusty plasma with an expanding background. The equations are
namely, the equations of continuity, momentum, and energy conservation,\begin{eqnarray}
\frac{\partial n_{j}}{\partial t}+\nabla\cdot(n_{j}\bm{v}_{j}) & = & 0,\\
m_{j}n_{j}\left(\frac{\partial\bm{v}_{j}}{\partial t}+\bm{v}_{j}\cdot\nabla\bm{v}_{j}\right) & = & -\nabla p_{j}-en_{j}\nabla\phi-m_{j}n_{j}\nabla\psi,\label{eq:momentum}\\
\frac{3}{2}n_{j}\frac{dT_{j}}{dt}+p_{j}\nabla\cdot\bm{v}_{j} & = & \chi_{j}\nabla^{2}T_{j}-m_{j}n_{j}{\cal L}_{j}(n_{j},T_{j}),\label{eq:energy}\end{eqnarray}
where the subscript $j=e$ and $d$ for the electrons and dust particles,
$\chi_{j}$ is the thermal conductivity of the species, and $\phi$
and $\psi$ are the electrostatic and gravitational potentials defined
by their respective Poisson's equations,\begin{eqnarray}
\nabla^{2}\phi & = & -4\pi[e(n_{i}-n_{e})+Qn_{d}],\\
\nabla^{2}\psi & = & 4\pi Gm_{d}n_{d},\end{eqnarray}
$Q$ being the total charge on the dust particles. The ratio of specific
heats $\gamma=5/3$ in these equations and other symbols have their
usual meanings. As mentioned before, we have assumed the massive dust
particle to be negatively charged. The second term on the right hand
side of Eq.(\ref{eq:energy}) represents the net heat-loss through
radiation \cite{fi-1}. Note that, in writing these equations, we
have assumed that the ions are Boltzmanian,\begin{equation}
n_{i}=n_{0}e^{-\frac{e\phi}{T_{i}}}.\label{eq:boltzmanian}\end{equation}
This assumption of thermalization of ions comes from the fact that
at temperature above $\sim10^{4}\,^{\circ}{\rm K}$, it is effectively
the electron energy, which gets radiated and the radiative cooling
of the electrons prevents them from attaining thermal equilibrium
\cite{bora,pandey}. Besides, as we are interested in the region of
perturbation with phase velocity of the order of $\sim\sqrt{T_{e}/m_{d}}$;
the ion thermal velocity ($\sim\sqrt{T_{e}/m_{i}}$) can be assumed
to be much higher, so that any temperature fluctuation in the ions
are assumed to be quickly equilibrated. The pressures for the electrons
and {\small dust} particles are given by $p_{j}=n_{j}T_{j}$.  In
what follows, we however neglect self-gravity, as the regimes where
our analysis might be applicable are not sufficiently massive to be
affected by self-gravity \cite{planetary}.

\subsection{Background (equilibrium) expansion with net cooling}

We consider now a homogeneous and uniformly expanding background and
assume that self-gravity is too weak to affect the expanding plasma
so that the equilibrium is characterized by an expansion parameter
$\varepsilon(t)$ \cite{wein,go-1},\begin{equation}
\bm{r}=\varepsilon(t)\bm{x},\quad\dot{\varepsilon}(t)=\mathrm{const.}\end{equation}
where $\bm{x}$ is the Lagrangian coordinate. Due to this non-static
background, the equilibrium quantities naturally become time dependent.
Note that when self-gravity is not negligible, the rate of expansion
is not constant and is given by,\begin{equation}
\ddot{\varepsilon}=-\frac{4}{3}\pi G\rho_{0}\varepsilon,\end{equation}
where $\rho_{0}$ is equilibrium matter density. The equilibrium background
velocity $\bm{v}_{0}$ is given by\begin{equation}
\bm{v}_{0}(\bm{r},t)=\frac{\dot{\varepsilon}}{\varepsilon}\bm{r}.\end{equation}
This expanding equilibrium is  mainly characterized by the equilibrium
density $n_{j0}$ and temperature $T_{j0}$,\begin{eqnarray}
\frac{dn_{j0}}{dt} & = & -3\frac{\dot{\varepsilon}}{\varepsilon}n_{j0},\\
\frac{d}{dt}\log T_{j0} & = & -2\left(\frac{\dot{\varepsilon}}{\varepsilon}+\frac{1}{3}m_{j}\frac{\mathcal{L}_{j0}}{T_{j0}}\right),\label{eq:equil_temp}\end{eqnarray}
where the equilibrium quantities are denoted by a subscript `0'. In
the expression for evolution of equilibrium temperature Eq.(\ref{eq:equil_temp}),
there are essentially two effects causing temperature to change ---
equilibrium background expansion {[}the first term in Eq.(\ref{eq:equil_temp})]
and net equilibrium cooling (the second term).

\section{The linear perturbation analysis}

We perturb the equilibrium with a small electrostatic perturbation
and for any arbitrary physical quantity $f(\bm{x},t)$, we write the
corresponding perturbation as\begin{equation}
f_{1}(\bm{x},t)\sim f_{1}(t)e^{i\bm{k}\cdot\bm{x}},\end{equation}
where $\bm{k}$ is the Lagrangian wave vector. In doing so, we have
retained the time dependent part of the perturbation exclusively with
the amplitude and Fourier decomposed the space part \cite{wein}.
We  split the perturbed velocity into two components, namely parallel
and perpendicular to $\bm{k}$,\begin{equation}
\bm{v}_{j1}(t)=\frac{\bm{k}}{k}v_{j1}(t)+\bm{v}_{j1\perp}(t),\end{equation}
where $v_{j1}(t)=\bm{\hat{k}}\cdot\bm{v}_{j1}$.

From the electron and dust continuity equations, the perturbed densities
can be written as,\begin{equation}
\frac{d\hat{n}_{j}}{dt}=-i\tilde{k}v_{j1},\label{eq:pert_continuity}\end{equation}
where the quantities with a `$\ \hat{}\ $' are perturbed quantities
normalized by their equilibrium values and $\tilde{k}=k/\varepsilon$.
If we neglect the inertia term of the electrons in the momentum equation,
Eq.(\ref{eq:momentum}), compared to the massive dust particles, the
perturbed electron pressure is just the perturbed electrostatic potential,\begin{equation}
\hat{p}_{e1}=\frac{e\phi_{1}}{T_{e0}}=\hat{\phi},\label{eq:pert_pressure}\end{equation}
in which, we have normalized the potential with the equilibrium electron
temperature. Note that the temperature is expressed in energy units.
As the ions are assumed to be Boltzmanian {[}Eq.(\ref{eq:boltzmanian})],
the perturbed ion density can simply be expressed as,\begin{equation}
\hat{n}_{i}=-\tau\hat{\phi},\label{eq:pert_ion_density}\end{equation}
where $\tau=T_{e0}/T_{io}$ is the ratio of the electron and ion equilibrium
temperatures.

From the electron energy equation Eq.(\ref{eq:energy}), we have,\begin{eqnarray}
\left(\frac{3}{2}n_{e}\frac{dT_{e}}{dt}+p_{e}\nabla\cdot\bm{v}_{e}\right)_{1}+\hat{p}_{e}(\chi_{0}\tilde{k}^{2}T_{e0}+m_{e}n_{e0}{\cal L}_{e0}+m_{e}n_{e0}T_{e0}^{2}{\cal L}_{n_{e}} & = & 0,\nonumber \\
-\hat{n}_{e}(\chi_{0}\tilde{k}^{2}T_{e0}+m_{e}n_{e0}T_{e0}^{2}{\cal L}_{p_{e}})\label{eq:energy1}\end{eqnarray}
where the $()_{1}$ denotes the first order perturbation of the respective
quantity and the equilibrium derivatives of the radiative loss function
are defined as,\begin{eqnarray}
{\cal L}_{n_{e}} & = & \left[\frac{\partial}{\partial T_{e0}}\left(\frac{{\cal L}_{e0}}{T_{e0}}\right)\right]_{n_{e0}},\\
{\cal L}_{p_{e}} & = & \left[\frac{\partial}{\partial T_{e0}}\left(\frac{{\cal L}_{e0}}{T_{e0}}\right)\right]_{p_{e0}}.\end{eqnarray}
The quantities inside the $()_{1}$ are given as,\begin{eqnarray}
\left(p_{e}\nabla\cdot\bm{v}_{e}\right)_{1} & = & p_{e0}(\nabla\cdot\bm{v}_{e1})-\hat{p}_{e}\left[\frac{3}{2}n_{e0}\frac{dT_{e0}}{dt}+m_{e}n_{e0}{\cal L}_{e0}\right],\\
\left(n_{e}\frac{dT_{e}}{dt}\right)_{1} & = & n_{e0}\frac{dT_{e1}}{dt}+n_{e0}(\hat{p}_{e}-\hat{T}_{e})\frac{dT_{e0}}{dt}.\end{eqnarray}
Finally using the perturbed electron continuity equation, Eq(\ref{eq:pert_continuity}),
we can write Eq.(\ref{eq:energy1}) for the perturbed electron temperature
as,\begin{equation}
\frac{d\hat{p}_{e}}{dt}=-i\frac{5}{3}\tilde{k}v_{e1}-\hat{p}_{e}(\tau_{\chi}^{-1}+\omega_{n})+\hat{n}_{e}(\tau_{\chi}^{-1}+\omega_{p}),\label{eq:pert_temp}\end{equation}
where $\tau_{\chi}$, $\omega_{n}^{-1}$, and $\omega_{p}^{-1}$ are
the characteristic time-scales for the thermal conduction by the electrons,
isochoric, and isobaric electron perturbations \cite{go-1}, given
by\begin{equation}
\tau_{\chi}^{-1}=\frac{2}{3}\frac{\chi_{0}\tilde{k}^{2}}{n_{e0}},\qquad\omega_{n,p}=\frac{2}{3}m_{e}T_{e0}{\cal L}_{n_{e},p_{e}}.\end{equation}

The quasi-neutrality condition with negatively charged dust particles
is given by,\begin{equation}
\hat{n}_{e}\delta_{e}-\hat{n}_{i}+(\hat{n}_{d}+\hat{Z}_{d})\delta_{d}=0,\label{eq:quasineutrality}\end{equation}
where we have taken the possibility of finite dust-charge fluctuation.
The quantities $\delta_{e}=n_{e0}/n_{i0}$ and $\delta_{d}=n_{d0}Z_{d0}/n_{i0}$
are the ratios of equilibrium electron and dust densities to the ion
densities and $Q=qZ_{d}$ is the dust charge with $Z_{d}$ being the
dust charge number. The dust charging equation is given by \cite{jana},\begin{equation}
\frac{dQ_{d1}}{dt}=I_{e1}+I_{i1},\end{equation}
where, $I_{j1}$s are perturbed electron and ion currents. The above
charging equation can be ultimately reduced to,\begin{equation}
\frac{dQ_{d1}}{dt}+\eta Q_{d1}=|I_{e0}|\left(\frac{n_{i1}}{n_{i0}}-\frac{n_{e1}}{n_{e0}}\right),\label{eq:pert_dust_charging}\end{equation}
where $\eta$ is natural decay rate of charge fluctuations \cite{jana}.

Finally, we write the dust momentum equation Eq.(\ref{eq:momentum})
as,\begin{equation}
\frac{dv_{d1}}{dt}=-\frac{\dot{\varepsilon}}{\varepsilon}v_{d1}+iZ_{d0}\omega_{d}c_{d}\phi_{1},\label{eq:pert_dust_momentum}\end{equation}
where \begin{equation}
\omega_{d}=\tilde{k}c_{d}\end{equation}
is the frequency of acoustic perturbation, and $c_{d}=\sqrt{{T_{e0}/m_{d}}}$.
By virtue of the non-static equilibrium, all the $\omega$s i.e. $\omega_{n,p,d}$
have become time dependent quantities.

The complete closed set of equations for linear perturbation of our
model is given by Eqs.(\ref{eq:pert_continuity})--(\ref{eq:pert_ion_density}),
(\ref{eq:pert_temp}), (\ref{eq:quasineutrality}), (\ref{eq:pert_dust_charging}),
and (\ref{eq:pert_dust_momentum}).

\subsection{Analysis of the perturbed system}

As the linearised differential equations become non-autonomous due
to equilibrium expansion and net cooling, the problem, in general,
is not amenable to standard Fourier decomposition for obtaining the
dispersion relation, which requires rigorous numerical treatment as
exact analytical solutions do not exist. However, depending on the
time-scales of the problem, we can apply approximation techniques
viz. WKB approximation \cite{bender,go-1} to gain valuable insight
into the nature of the problem. Recently Nejad-Asghar and Ghanbari \cite{nejad}
have considered the problem of non-static equilibrium for radiation
instability in molecular clouds, where they have used an ansatz of
exponential form for the normalized perturbed quantities. In our notation,
their formalism can be expressed as,\begin{equation}
\hat{n}_{d}\sim\frac{1}{\varepsilon(t)}e^{\omega t},\label{eq:exp_decomp}\end{equation}
where the quantity $\omega$ is analogous to normal mode frequency,
which is decoupled from $t$. However, as the equilibrium is time
dependent, the normal modes of the system become time dependent too,
so that $\omega=\omega(t)$ and a decomposition of the kind expressed
in Eq.(\ref{eq:exp_decomp}) is not possible. As we show in the appendix,
perturbation amplitude of densities based on Eq.(\ref{eq:exp_decomp})
does not lead too far and grossly disagrees with the numerical analysis
and the validity of the corresponding results are very limited.

In the following Section, we use the WKB approximation to analyse
the perturbed system in the limit of static dust-charge.

\section{Static dust charge : \emph{WKB approximation}}

In order to simplify the mathematical complexity of the problem, we
first consider the case of static dust charge. In this limit, we can
reduce the relevant equations of Sec.II to form a third order differential
equation in any of the perturbed variables, which can be conveniently
expressed as,\begin{equation}
a_{3}(t)\frac{d^{3}\hat{n}_{d}}{dt^{3}}+a_{2}(t)\frac{d^{2}\hat{n}_{d}}{dt^{2}}+a_{1}(t)\frac{d\hat{n}_{d}}{dt}+a_{0}(t)\hat{n}_{d}=0,\label{eq:diff_eq}\end{equation}
where the coefficients $a_{i}(t)$s are given by,\begin{eqnarray}
a_{3}(t) & = & 3\delta_{e}+5\tau(t)\label{eq:a3}\\
a_{2}(t) & = & 3\delta_{e}\Omega_{n}+3\tau(t)\Omega_{p}+6\delta_{e}\omega_{c}+2\omega_{\varepsilon}[9\delta_{e}+10\tau(t)],\\
a_{1}(t) & = & 2a_{2}(t)-\omega_{\varepsilon}[9\delta_{e}+10\tau(t)]+5\delta_{d}Z_{d0}(t)\omega_{d}^{2},\\
a_{0}(t) & = & 3\delta_{d}Z_{d0}(t)\omega_{d}^{2}\Omega_{p},\label{eq:a0}\end{eqnarray}
Various frequencies used in Eq.(\ref{eq:diff_eq}) are defined as\begin{eqnarray}
\Omega_{n} & = & \omega_{n}(t)+\tau_{\chi}^{-1}(t),\\
\Omega_{p} & = & \omega_{p}(t)+\tau_{\chi}^{-1}(t).\end{eqnarray}
The expansion and cooling time-scales are given by $\omega_{\varepsilon,c}^{-1}$,
\begin{equation}
\omega_{\varepsilon}=\left(\frac{\dot{\varepsilon}}{\varepsilon}\right),\qquad\omega_{c}=\left(\frac{1}{3}m_{e}\frac{{\cal L}_{e0}}{T_{e0}}\right).\end{equation}
Different time-scales of the system are dictated by $\omega_{n,p,c,\varepsilon,\chi}^{-1}$,
a clear separation of which is essential for WKB approximation.

\subsection{Linear dispersion relation}

Going by the standard WKB theory \cite{bender}, we expand the eigenfunctions
$\hat{n}_{d}(t)$ of Eq.(\ref{eq:diff_eq}) as,\begin{equation}
\hat{n}_{d}(t)\sim\exp\left[\frac{1}{\delta}\sum_{n=0}^{\infty}\delta^{n}S_{n}(t)\right],\quad\delta\rightarrow0,\end{equation}
where $\delta$ is a small parameter and $S_{i}(t)$ are arbitrary
functions of time. We break up the coefficients $a_{i}(t)$s in time
as,\begin{equation}
a_{i}(t)\equiv a_{i}^{(0)}+a_{i}^{(1)}\epsilon^{-1}+a_{i}^{(2)}\epsilon^{-2}+\cdots,\label{eq:expansion}\end{equation}
where $\epsilon$ is the smallest time-scale of the system. We assume
\emph{a-priori} that the different coefficients $a_{i}(t)$s can be
ordered as,\begin{eqnarray}
a_{3}(t) & = & a_{3}^{(0)},\\
a_{2}(t) & = & a_{2}^{(0)}+a_{2}^{(1)}\epsilon^{-1},\\
a_{1}(t) & = & a_{1}^{(0)}+a_{1}^{(1)}\epsilon^{-1}+a_{1}^{(2)}\epsilon^{-2},\\
a_{0}(t) & = & a_{0}^{(0)}+a_{0}^{(1)}\epsilon^{-1}+a_{0}^{(2)}\epsilon^{-2}+a_{0}^{(3)}\epsilon^{-3}.\end{eqnarray}
Assuming $\epsilon\sim\delta=\epsilon$, we substitute the above approximations
along with expansion given by Eq.(\ref{eq:expansion}) in Eq.(\ref{eq:diff_eq}).
After equating the like powers of $\epsilon$ in the resultant relation,
we obtain the linear dispersion relation for the system, from Eq.(\ref{eq:diff_eq})
as,\begin{eqnarray}
a_{3}^{(0)}\left(3\sum_{j=0}^{N-1}S_{j}^{\prime}S_{N-1-j}^{\prime\prime}+S_{N-2}^{\prime\prime\prime}+\sum_{j=0}^{N}\sum_{l=0}^{N-j}S_{j}^{\prime}S_{l}^{\prime}S_{N-l-j}^{\prime}\right) & = & 0,\quad N=0,1,\dots,\infty\label{eq:wkb_dr}\\
+a_{2}^{(0)}\left(\sum_{j=0}^{N-1}S_{j}^{\prime}S_{N-1-j}^{\prime}+S_{N-2}^{\prime\prime}\right)+a_{2}^{(1)}\left(\sum_{j=0}^{N}S_{j}^{\prime}S_{N-j}^{\prime}+S_{N-1}^{\prime\prime}\right)\nonumber \\
+a_{1}^{(0)}S_{N-2}^{\prime}+a_{1}^{(1)}S_{N-1}^{\prime}+a_{1}^{(2)}S_{N}^{\prime}+a_{0}^{(3-N)}\nonumber \end{eqnarray}
where\begin{eqnarray}
a_{i}^{(j)},\ S_{j}=0, & \mathrm{for} & j<0,\\
 &  & i=0,1,2,3\nonumber \end{eqnarray}
We note that apart from the basic assumption of well separated time-scales
of our problem, two essential conditions for validity of the WKB expansion
are, as $\epsilon\rightarrow0$, \begin{eqnarray}
\left|\epsilon^{n}S_{n+1}(t)\right| & \ll & \left|\epsilon^{n-1}S_{n}(t)\right|,\\
\left|\epsilon^{N}S_{N+1}(t)\right| & \ll & 1.\end{eqnarray}

We express the WKB function $S_{n}(t)$ as an integral of an unknown
function $\psi_{n}(t)$,\begin{equation}
S_{n}(t)\sim\int^{t}\tau_{0}^{n-1}\psi_{n}(t)\, dt,\end{equation}
where $\tau_{0}$ is the shortest time-scale associated with the physical
system. The WKB dispersion relation, Eq.(\ref{eq:wkb_dr}) can now
be examined in different regimes depending on their characteristic
time-scales.

\subsection{The dust-acoustic domain ($\omega_{c}\sim\omega_{\varepsilon}\sim\omega_{p,n}\sim\tau_{\chi}^{-1}\ll\omega_{d}$)}

In the dust-acoustic domain, we assume that the characteristic time-scale
is dictated by the dust-sound frequency $\omega_{d}$ which is the
dominant one. It should however be noted that $\omega_{d}$ is not
the usual dust-acoustic frequency which can be written as $\sqrt{T_{d}/m_{d}}$.
The WKB coefficients \cite{go-1} $a_{i}^{(j)}$ can now be ordered
as $a_{3}=a_{3}^{(0)},a_{2}=a_{2}^{(0)},a_{1}=a_{1}^{(2)}+a_{1}^{(0)},$
and $a_{0}=a_{0}^{(2)}$. The second order coefficients are given
by\begin{eqnarray}
a_{1}^{(2)} & = & 5\delta_{d}Z_{d0}\omega_{d}^{2},\\
a_{0}^{(2)} & = & 3\delta_{d}Z_{d0}\omega_{d}^{2}\Omega_{p}.\end{eqnarray}
 To the zeroth order ($N=0$), the dispersion relation Eq.(\ref{eq:wkb_dr})
now becomes\begin{equation}
(3\delta_{e}+5\tau(t))(\omega_{0}\psi_{0})^{3}+5\delta_{d}Z_{d0}\omega_{d}^{2}(t)(\omega_{0}\psi_{0})=0,\end{equation}
which has two solutions for the condensation and acoustic modes respectively,\begin{eqnarray}
\psi_{0{\rm c}} & = & 0,\\
\omega_{0}\psi_{0{\rm s}} & = & \pm i\sqrt{{\frac{a_{1}^{(2)}(t)}{a_{3}^{(0)}(t)}}}=\pm i\left(\frac{5\delta_{d}Z_{d0}\omega_{d}^{2}(t)}{3\delta_{e}+5\tau(t)}\right)^{1/2},\label{eq:wkb_sound}\end{eqnarray}
where $\omega_{0}=\tau_{0}^{-1}$ and the subscripts `c' and `s' refer
to the condensation and sound modes.

The first order contribution ($N=1$) to the condensation mode is
given by\begin{equation}
\psi_{1{\rm c}}=-\frac{3}{5}\Omega_{p}.\label{eq:wkb_condensation}\end{equation}
So, the condition for a growing condensation mode is given by the
condition $\Omega_{p}<0$, \begin{equation}
\frac{\chi_{0}\tilde{k}^{2}}{n_{e0}}+m_{e}T_{e0}\frac{\partial}{\partial T_{e0}}\left(\frac{{\cal L}_{e0}}{T_{e0}}\right)_{p_{e0}}<0,\label{eq:cond_inst_condition}\end{equation}
which is similar to the condition for a static equilibrium \cite{fi-1,go-1}.
As can be seen from the above condition (\ref{eq:cond_inst_condition}),
thermal conductivity has a stabilising effect on temperature perturbation.
The corresponding critical wave number for the condensation mode is
given by \cite{go-1},\begin{equation}
k_{c}^{2}=-\varepsilon^{2}T_{e0}\frac{m_{e}n_{e0}}{\chi_{0}}\frac{\partial}{\partial T_{e0}}\left(\frac{{\cal L}_{e0}}{T_{e0}}\right)_{p_{e0}},\label{eq:k_crit}\end{equation}
beyond which the condensation mode is stabilized. Note that the critical
wave number has become dependent on the equilibrium electron density,
which means that as more and more electrons get depleted on the surface
of the massive dust particles, perturbation with longer wavelengths
becomes stable \cite{bora}. This is obvious as the cooling is due
to the electrons, the depletion of electrons by the dust particles
prevents localised regions inside a diffuse structure from effectively
radiating. The dust can, however radiate at a much longer wavelength
to which the medium may be opaque \cite{ol-1,ba-1}. Comprehensive
effect of the presence of dust particles can only be realised when
we consider the effect of dust-charge fluctuation in Section.V.

The first order contribution to the sound mode can be obtained from
the corresponding dispersion relation as\begin{equation}
3a_{3}^{(0)}\omega_{0}^{2}(\psi_{0s}\dot{\psi}_{0s}+\psi_{1s}\psi_{0s}^{2})+a_{2}^{(0)}(\omega_{0}\psi_{0s})^{2}+a_{1}^{(2)}\psi_{1s}+a_{0}^{(2)}=0,\end{equation}
which can be solved for $\psi_{1s}$,\begin{equation}
\psi_{1s}=\frac{3}{2a_{3}^{(0)}}\left[a_{3}^{(0)}(2\omega_{\varepsilon}+\omega_{c})-\tau(\omega_{\varepsilon}+\omega_{c})\right]-\frac{a_{2}^{(0)}}{2a_{3}^{(0)}}+\frac{a_{0}^{(2)}}{2a_{1}^{(2)}},\label{eq:wkb_sound}\end{equation}
which is real. So the existence of a growing sound mode (overstable)
at $t=0$ depends on whether $\psi_{1s}>0$ at $t=0$.

\subsubsection{Validity of WKB approximation}

\begin{figure}[t]
\begin{centering}
\includegraphics[width=0.45\textwidth]{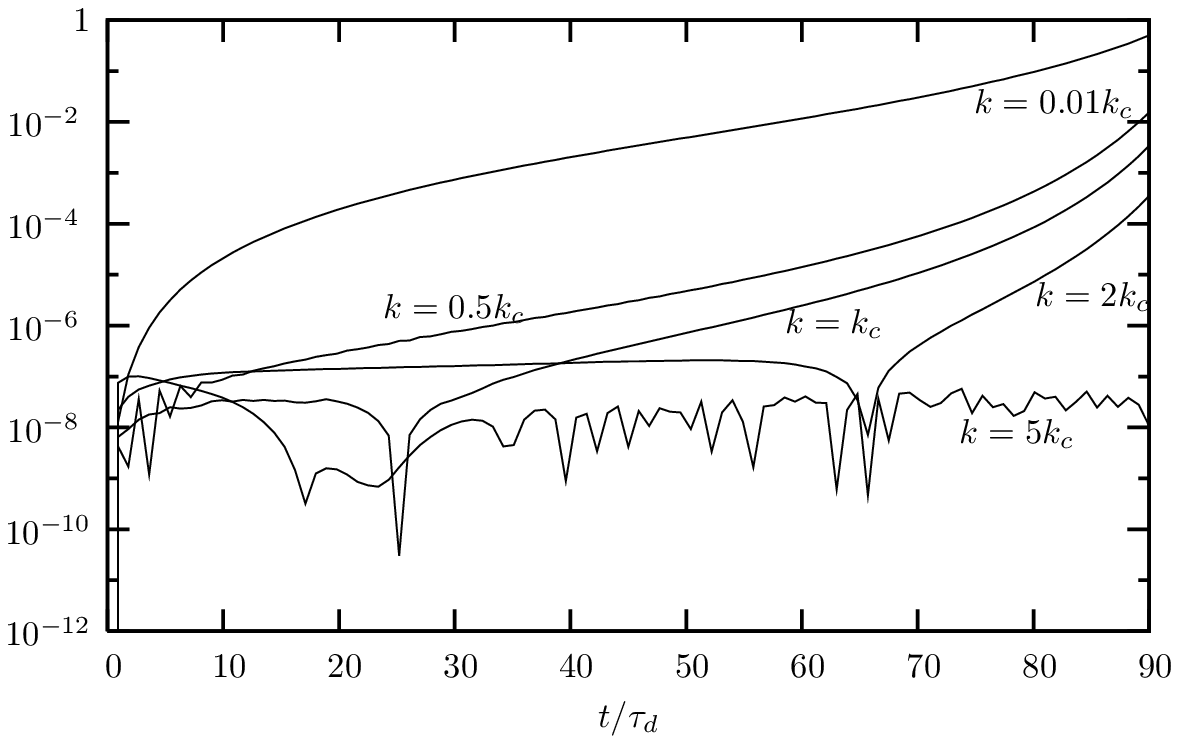}\hfill{}\includegraphics[width=0.45\textwidth]{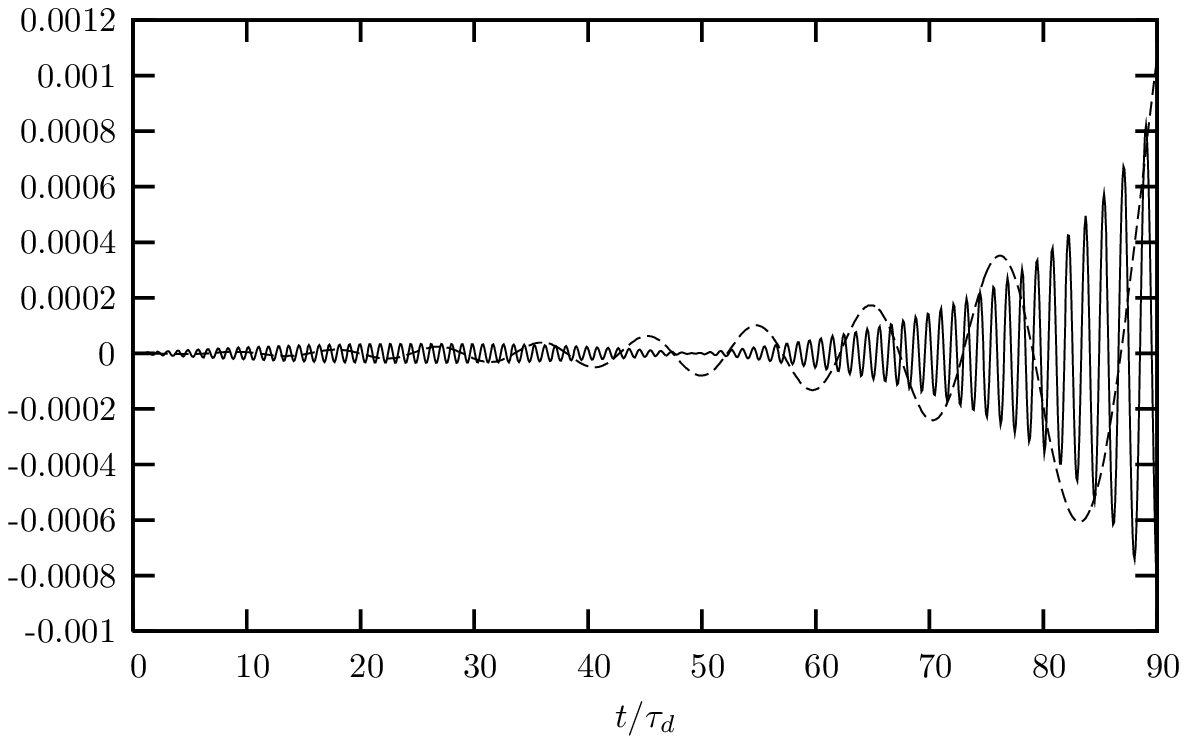}
\par\end{centering}

\caption{\label{cap:wkb_error}Comparison between the WKB and numerically
computed solutions of Eq.(\ref{eq:diff_eq}) for a static equilibrium.
Relative errors for different wave numbers for the condensation mode
are shown in the first figure. In the second figure, relative errors
for the sound mode for wave numbers $k_{0}=0.1k_{c}$ (dashed) and
$k_{0}=k_{c}$ (solid) are shown. The parameters $\alpha_{c}=0.01$,
$\beta_{\chi}=0.5$, and $\tau_{d}=\omega_{d}^{-1}$. }
\end{figure}

At this point, we would like to compare the relative errors between
the WKB solutions, namely (\ref{eq:wkb_condensation}) and (\ref{eq:wkb_sound})
of the condensation and the sound modes, and the numerically computed
solutions of Eq.(\ref{eq:diff_eq}). As can be seen from Fig.\ref{cap:wkb_error},
in the absence of expansion, the relative deviation of the WKB solutions
from the numerically computed solutions is within $10^{-4}$ except
for highly unstable perturbations of very long wavelengths, $k\ll k_{c}$.
For short wavelength perturbation, $k>k_{c}$, the deviation is down
to $10^{-8}$.%

In the above analysis, we have assumed that the equilibrium heat-loss
function \cite{spitzer} $\mathcal{L}_{e0}\propto n_{e0}T_{e0}^{1/2}$
so that the critical wave number $k_{c}$ can be expressed as\begin{equation}
\frac{k_{c}}{k_{0}}=\varepsilon\left(\frac{3\alpha_{c}}{\beta_{\chi}}\right)^{1/2},\label{eq:critical_waveno}\end{equation}
where $k_{0}=\tilde{k}_{t=0}$. In the above expression, $\alpha_{c}$
and $\beta_{\chi}$ denote the ratios of the characteristic equilibrium
cooling time and the thermal conduction time to the dust-acoustic
time at the beginning of the expansion ($t=0$), \begin{equation}
\alpha_{c}=\frac{\omega_{c0}}{\omega_{0}},\qquad\beta_{\chi}=\frac{\tau_{\chi0}^{-1}}{\omega_{0}},\end{equation}
where $\left(\omega_{c0},\tau_{\chi0}^{-1}\right)=\left(\omega_{c},\tau_{\chi}^{-1}\right)_{t=0}$
and $\omega_{0}$ is the equilibrium dust-acoustic frequency at $t=0$,
\[
\omega_{0}^{-1}=\frac{1}{k_{0}}\sqrt{\frac{m_{d}}{T_{0}}},\]
 $T_{0}$ being the initial equilibrium electron temperature $[T_{e0}]_{t=0}$.
We would like to point out that the present form of heat-loss function
inherently implies net equilibrium cooling as there is no heating
source. Any non-zero equilibrium value of $\mathcal{L}_{e0}$ actually
determines the net cooling.

In working out the WKB solutions we have assumed massive dust particles
\cite{mendis} with mass of the order of $10^{18}m_{H}$, $m_{H}$
being the proton mass and the dust charge number $Z_{d0}=10^{3}$.
The electron to dust ratio is fixed at $\delta_{e}=0.1$ i.e. a ninety
percent depletion of the electrons on the surface of the dust particles.
The initial electron temperature is assumed to be $T_{e0}=10^{7}\,^{0}{\rm K}$
at $t=0$. The initial equilibrium electron density $n_{e0}\sim10^{4}\,{\rm {\rm cm}}^{-3}$
is taken to be of the order of the average electron density in multiple-shell
PNe (MSPNe) \cite{guer-1,stran-1}. The electron thermal conductivity
at this temperature can be written as \cite{rybicki,spitzer},\begin{equation}
\chi_{0}=\frac{1.84\times10^{-5}}{\log\Lambda_{c}}T_{0}^{5/2}\,{\rm ergs\, s}^{-1}\,{\rm K}^{-1}\,{\rm cm}^{-1},\end{equation}
where $\log\Lambda_{c}\approx30$ is the Coulomb logarithm. The domain
of instability is defined as\begin{equation}
\omega_{c0}<\tau_{\chi0}^{-1}<\omega_{0},\end{equation}
as thermal conductivity is more effective at short scale length. We
have assumed a value of $\alpha_{c}=0.01$ and $\beta_{\chi}=0.5$,
for which the critical wave number, given by Eq.(\ref{eq:critical_waveno}),
is $\sim2.7\times10^{-10}\,{\rm cm}^{-1}$ corresponding to a critical
wavelength $\lambda_{c}\sim2.3\times10^{10}\,{\rm cm}\approx7.6\times10^{-9}\,{\rm pc}$.
As the average size of a PNe is of the order of $0.1\,{\rm pc}$ \cite{guer-1},
a perturbation wavelength of the order of a few thousand kilometers
($\sim\lambda_{c})$ is quite small compared to the size of the nebula.
The characteristic time-scale $\tau_{d}\sim10^{3}\,{\rm years}$,
which is about 10\% of the average kinematic age of a MSPNe \cite{chou-1}.

\begin{figure}[t]
\begin{centering}
\includegraphics[width=0.45\textwidth]{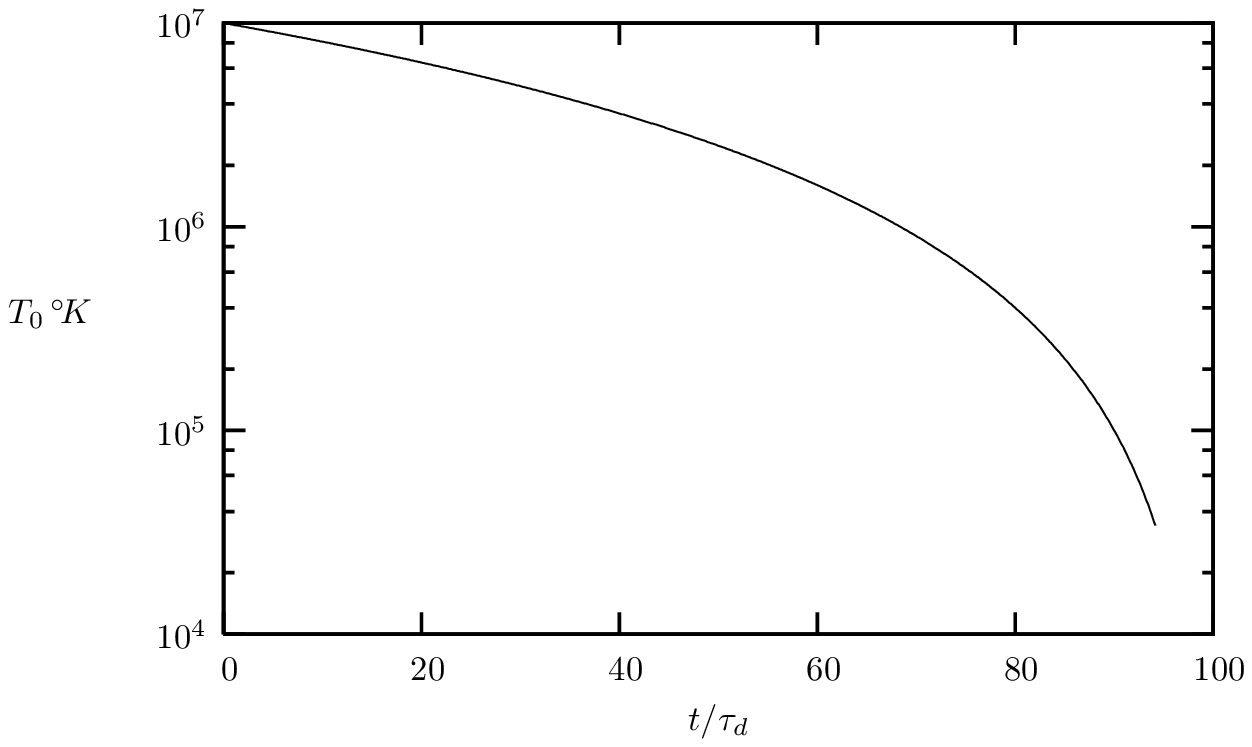}\hfill{}\includegraphics[width=0.47\textwidth]{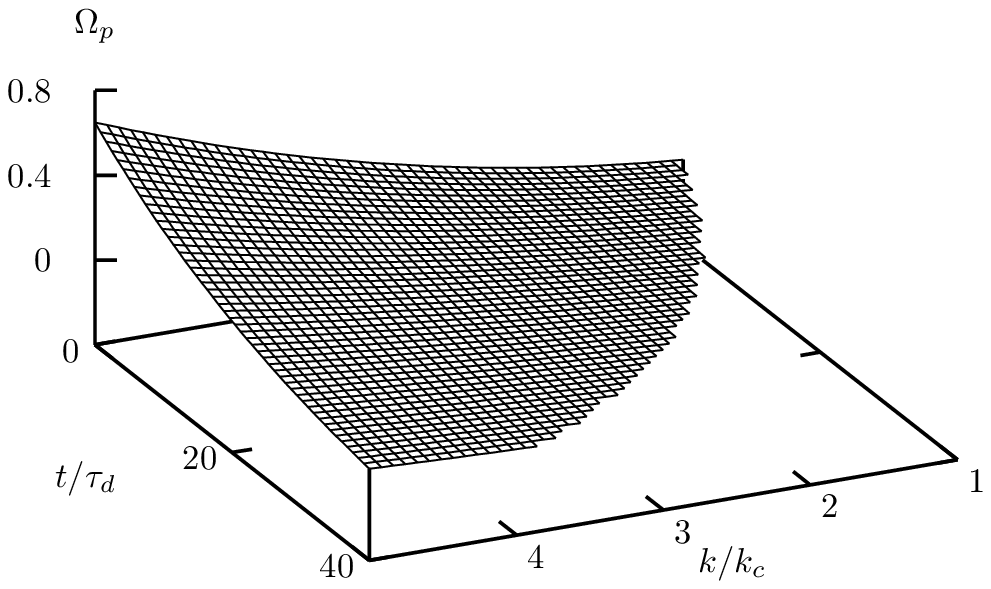}
\par\end{centering}

\caption{\label{cap:eq_T}Equilibrium evolution of electron temperature $T_{e0}(t)$
in absence of expansion (first diagram). The isobaric stability criterion
is plotted in the second diagram against $k/k_{c}$ and $t$. For
clarity, the negative values of $\Omega_{p}$ part of the $\Omega_{p}-k-t$
surface are hidden which signifies instability. $\Omega_{p}$ in the
figure is normalized by the initial dust-sound frequency $\omega_{0}$.}
\end{figure}

The equilibrium evolution of electron temperature can be determined
from Eq.(\ref{eq:equil_temp}), which in terms of the expansion parameter
$\varepsilon$ can be written as,\begin{equation}
T_{e0}(t)=T_{0}\left[\frac{1-(\omega_{c0}-\alpha)t}{(1+\alpha t)^{2}}\right]^{2},\qquad\omega_{c0}t\leq1,\label{eq:temp_evol}\end{equation}
where $\alpha=\dot{\varepsilon}$ is the rate of expansion, which
is constant in this case. The evolution of $T_{e0}(t)$ in the absence
of expansion is shown in Fig.\ref{cap:eq_T}. Note that relation (\ref{eq:temp_evol})
is valid as long as $t<\omega_{c0}^{-1}$. In general, equilibrium
cooling should vanish as temperature approaches zero. We note that
in absence of equilibrium cooling ($\omega_{c0}=0$), the time dependence
of the equilibrium temperature can be expressed as\begin{equation}
T_{e0}(t)=T_{0}\varepsilon(t)^{-3(\gamma-1)},\end{equation}
where $\gamma$ is the ratio of specific heat, which reduces to $T_{0}\varepsilon(t)^{-2}$
for $\gamma=5/3$ in our case. %

\subsubsection{Effect of equilibrium cooling and expansion}

Before we proceed further, a few comments and observations are to
be made. If we consider only a static equilibrium with net cooling,
any dependence of the equilibrium quantities is due to the evolution
of equilibrium temperature. We note that as the stability (or instability)
condition for the condensation mode i.e. condition (\ref{eq:cond_inst_condition})
is valid throughout the whole time domain, the expression for critical
wave number, given by Eq.(\ref{eq:k_crit}), is dynamic due to the
time dependence of the equilibrium quantities. The critical wave number
is dynamically going to change, effectively determined by the detailed
temperature dependence of the heat-loss function ${\cal L}_{e0}$.
So, if we start off initially with a long wavelength stable perturbation,
it might become unstable at a later time depending on the instantaneous
value of the critical wave number $k_{c}$. As an example, we have
shown the evolution of the quantity $\Omega_{p}$ in time with different
$k/k_{c}$, in the second diagram of Fig.\ref{cap:eq_T}, for the
parameter range mentioned later in this section. As can be seen from
the figure, even if the initial value for $\Omega_{p}$ at $t=0$
is positive (signifying stability for the condensation mode) for say
$k\sim2k_{c}$, at a later time, around $t\approx25\tau_{d}$, the
quantity falls below zero marking the onset of the instability. Similar
observations can also be made for the sound mode. For the form of
the heat-loss function $\mathcal{L}_{e0}\propto n_{e0}T_{e0}^{1/2}$,
the instability condition (\ref{eq:cond_inst_condition}) can be written
as,\begin{equation}
\beta_{\chi}\frac{[1-(\omega_{c0}-\alpha)t]^{5}}{(1+\alpha t)^{9}}k^{2}-3\omega_{c0}<0,\end{equation}
from which we see that both equilibrium cooling and expansion destabilize
the condensation mode and an otherwise stable mode ultimately may
become unstable due to either equilibrium cooling or expansion or
both. However, the behavior of the growth rate can be ascertained
only after a detailed calculation. Apparently, the instability condition
for the condensation mode does not depend upon the presence of dust
particles as it is just a balance between the electron conductivity
and cooling due to electrons, both of which are equally affected by
the dust particles.

As mentioned earlier, the existence of overstable mode depends on
the positivity of $\psi_{1s}$, given by Eq.(\ref{eq:wkb_sound}),
which can be simplified as,\begin{equation}
\psi_{1s}=\left(\frac{7}{2}\omega_{\varepsilon}+6\omega_{c}\right)\tau+\frac{3}{5}\left(\frac{1}{2}\omega_{c}-\tau_{\chi}^{-1}\right)\delta_{e}.\label{eq:wkb_sound_simple}\end{equation}
As $\delta_{e}\leq1$, we can write the condition of unstable acoustic
mode as\begin{equation}
\frac{3}{5}\tau_{\chi}^{-1}\delta_{e}<\tau\left(\frac{7}{2}\omega_{\varepsilon}+6\omega_{c}\right)\tau,\end{equation}
which, in the limit of vanishing background expansion ($\omega_{\varepsilon}=0$)
and $\tau\sim1$, becomes $\tau_{\chi}^{-1}\delta_{e}<10\omega_{c}$
and for $\tau\ll1$, reduces to $\omega_{c}>2\tau_{\chi}^{-1}$. So,
the presence of dust ($\delta_{e}<1$) actually destabilizes the acoustic
mode and a stable acoustic mode can become unstable due to the capturing
of the electrons by the massive dust particles. This is due to the
decrease in thermal conduction by the electrons, as more and more
electrons are captured by the massive dust particles. The relation,
Eq.(\ref{eq:wkb_sound_simple}) without background expansion ($\alpha=0$),
can be expressed as,\begin{equation}
\psi_{1s}=\frac{3\omega_{c0}}{(1-\omega_{c0}t)}\left(2\tau+\frac{1}{10}\delta_{e}\right)-\frac{3}{5}\beta_{\chi}(1-\omega_{c0}t)^{5}k^{2}\delta_{e},\end{equation}
which shows that equilibrium cooling destabilizes the acoustic mode.

The immediate effect of equilibrium expansion is that the normal modes
of the system becomes time dependent, which is analogous to change
in frequency of vibration in Doppler effects. However the stability
of a certain mode may depend on balancing between the opposing effects.
For example, the instability condition for the acoustic mode in the
limit of very small equilibrium cooling ($\omega_{c0}\ll1$), can
be expressed as,\begin{equation}
\frac{7\alpha\tau}{(1+\alpha t)}>\frac{6\beta_{\chi}k^{2}\delta_{e}}{(1+\alpha t)^{4}},\end{equation}
and we see that background expansion too destabilizes the acoustic
mode. Physically, expansion nullifies the effect of thermal conductivity
as thermal conductivity is most effective at short scale lengths,
leaving the onset of thermal instability entirely on the isobaric
cooling.

\begin{figure}[t]
\begin{centering}
\includegraphics[width=0.45\textwidth]{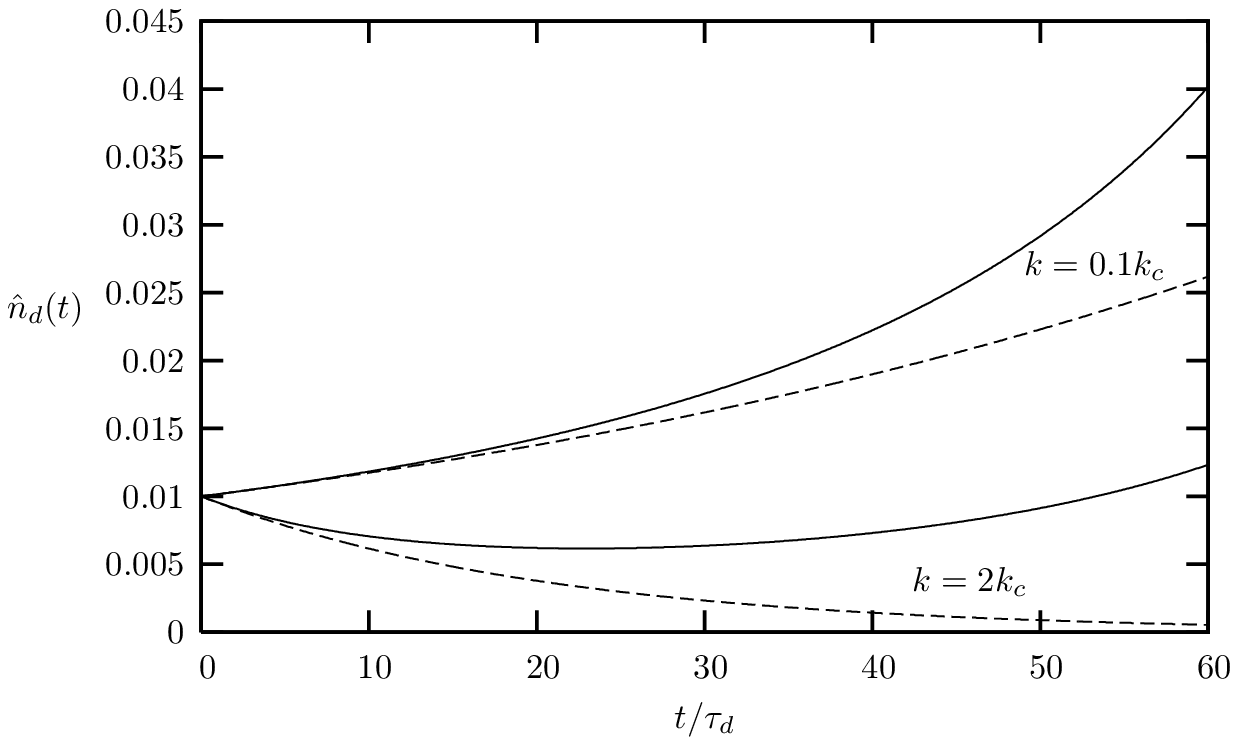}\hfill{}\includegraphics[width=0.45\textwidth]{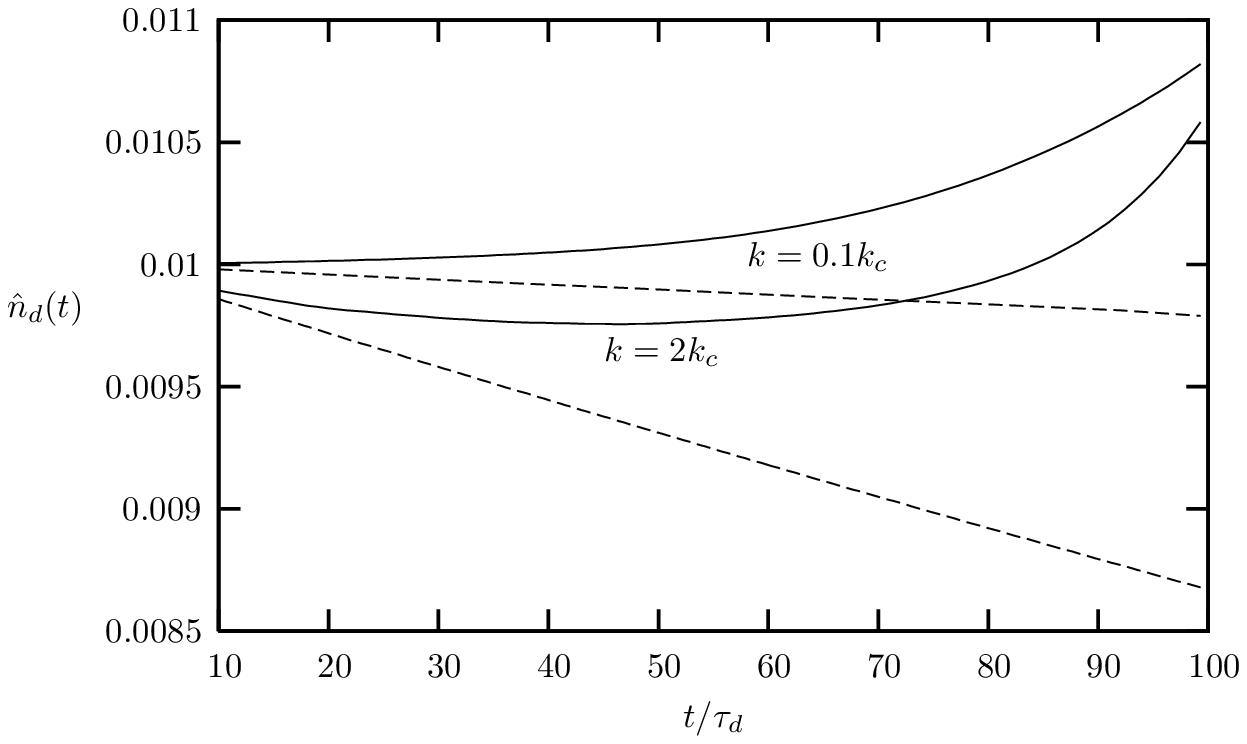}
\par\end{centering}

\caption{\label{cap:only_cooling}Comparison with the classical case \cite{fi-1}
for static equilibrium (no expansion) in presence of equilibrium temperature
evolution for the perturbed dust density for the condensation mode
(first figure) and wave mode (second figure). Rest of the parameters
are same as for Fig.\ref{cap:eq_T}. The dashed lines indicate the
classical Field's behavior. The values of wave number are shown is
at $t=0$.}
\end{figure}

The cumulative effect of equilibrium cooling, background expansion,
and electron capturing my massive dust particles is to destabilize
the condensation and the acoustic modes which effectively increases
(reduces) the cut-off wavenumber (wavelength) beyond which the modes
are stable. Therefore, we conclude that as the nebulae evolve due
to expansion, the microstructures eventually die out and the presence
of dust particles enhances this effect, which actually conforms to
the observational studies, which report that the microstructures are
transient compared to the evolutionary life (due to expansion) of
the planetary nebulae\cite{planetary}.

We now compute the WKB solutions of the normalized perturbed dust
density $\hat{n}_{d}(t)$ for the parameter range discussed above.
The evolution of the perturbed dust density with only equilibrium
cooling for the condensation mode and the acoustic mode are shown
in Fig.\ref{cap:only_cooling}. As can be seen from the figures, though
the classical condensation mode is stable beyond the critical wave
number $k_{c}$, in presence of equilibrium cooling, the condensation
mode can grow further. However, for $k\gg k_{c}$, the mode is stabilized.
The equilibrium cooling has similar effects on the acoustic mode (see
the second figure in Fig.\ref{cap:only_cooling}). The classical results
with no equilibrium expansion and cooling \cite{fi-1} are shown as
dashed curves. The initial level of perturbation is fixed at $0.01$
in all the cases.

\begin{figure}[t]
\begin{centering}
\includegraphics[width=0.45\textwidth]{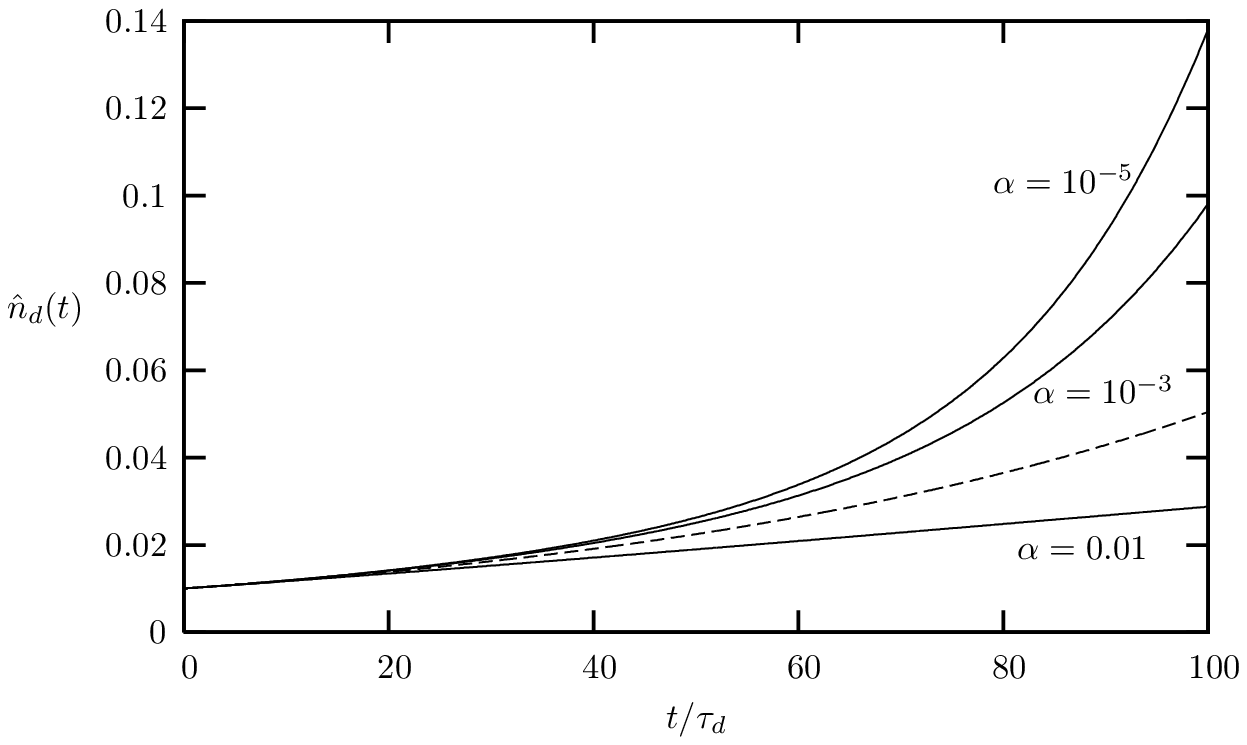}\hfill{}\includegraphics[width=0.45\textwidth]{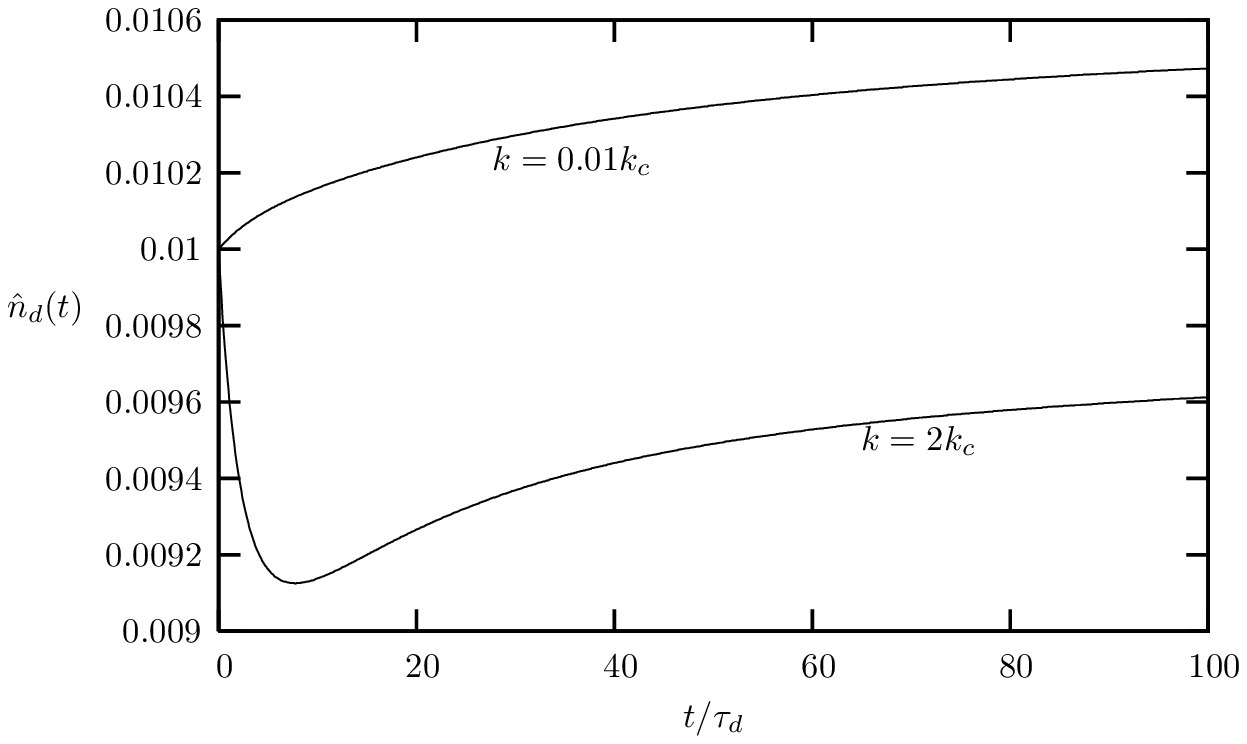}
\par\end{centering}

\caption{\label{cap:cooling_expansion}In the first diagram the comparison
of the WKB solutions for the condensation mode with the classical
Field's case in presence of equilibrium expansion and net equilibrium
cooling, is shown for $k=0.01\varepsilon k_{c}$. Other parameters
are same as before and the dotted line shows the prediction of the
classical theory. In the second diagram, the destabilisation of the
condensation mode for rapid expansion ($\alpha=0.1$) is shown for
different $k$.}
\end{figure}

In Fig.\ref{cap:cooling_expansion}, we show the effect of equilibrium
expansion on the modes. The rate of equilibrium expansion is given
by $\alpha$. As seen from the first diagram of Fig.\ref{cap:cooling_expansion},
rapid expansion causes the growth rate of the condensation mode to
decrease in the initial phase. However, finally, expansion destabilises
an otherwise stable mode, which is shown in the second diagram of
Fig.\ref{cap:cooling_expansion}, where we have set the expansion
rate at $\sim10\%$ of the dust-acoustic frequency.

\subsection{The heat-conduction domain ($\omega_{c},\omega_{\varepsilon},\omega_{p,n},\omega_{d}\ll\tau_{\chi}^{-1}$)}

At very small spatial scale, the heat conduction through the electrons
dominates and the thermal equilibration of the electrons prevents
any thermal instability. With this ordering, the WKB expansion yields
one condensation mode\begin{equation}
\omega_{0}\psi_{0{\rm c}}=-\frac{3\tau_{\chi}^{-1}(\tau+\delta_{e})}{(3\delta_{e}+5\tau)},\end{equation}
which is stable.

\section{Dust charge fluctuation : \emph{numerical solution}}

In this section we let the dust charge to change according to the
dust charging equation Eq.(\ref{eq:pert_dust_charging}). In what
follows, we however neglect the natural decay rate of charge fluctuations
$\eta$. Using Eqs.(\ref{eq:pert_ion_density}) and (\ref{eq:pert_pressure})
we can write the charging equation as\begin{equation}
\frac{d\hat{Z}_{d}}{dt}=\nu(-\tau\hat{p}_{e}-\hat{n}_{e}),\end{equation}
where $\hat{Z}_{d}$ is normalized perturbed dust-charge number and
$\nu$ is the charge fluctuation frequency normalized to initial dust-acoustic
frequency $\omega_{0}$. The fluctuation of dust charge introduces
another time-scale into the problem, which we would like to keep at
an arbitrary level. Therefore a clear separation of time-scales is
not possible and we investigate the dispersion relation numerically
instead of applying WKB approximation.

\begin{figure}[t]
\begin{centering}
\includegraphics[width=0.48\textwidth]{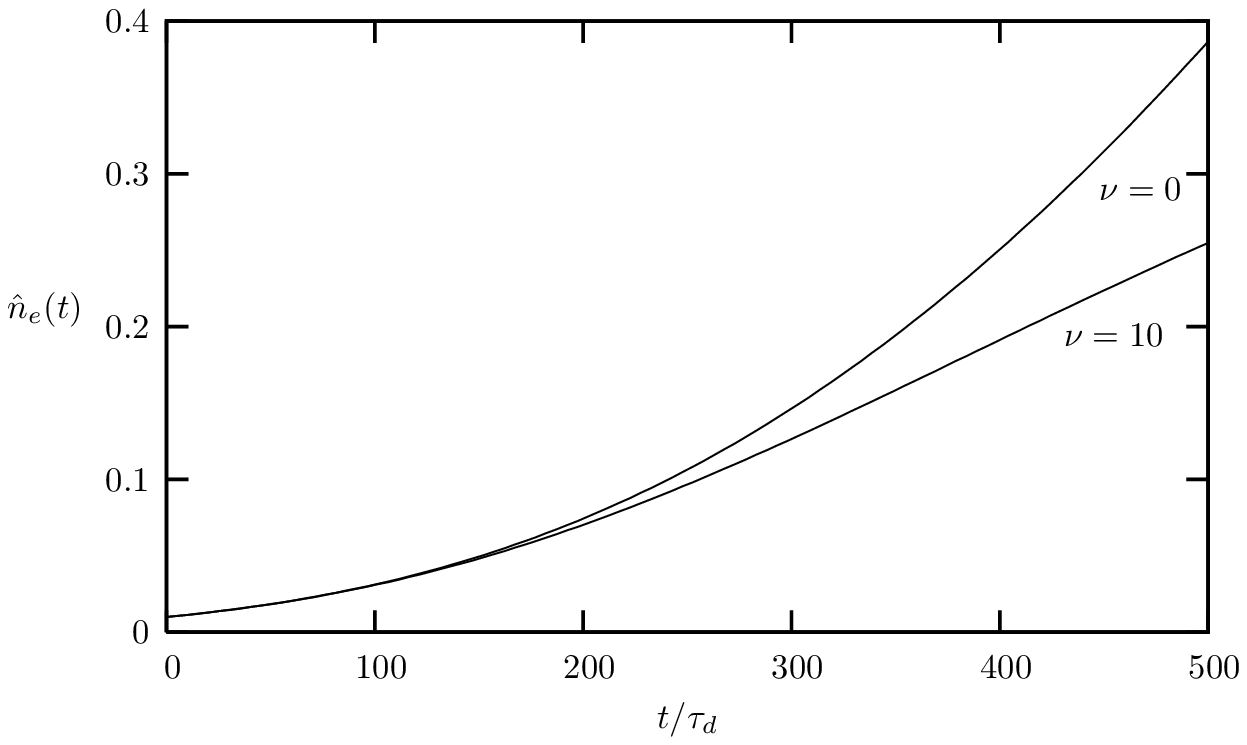}\hfill{}\includegraphics[width=0.48\textwidth]{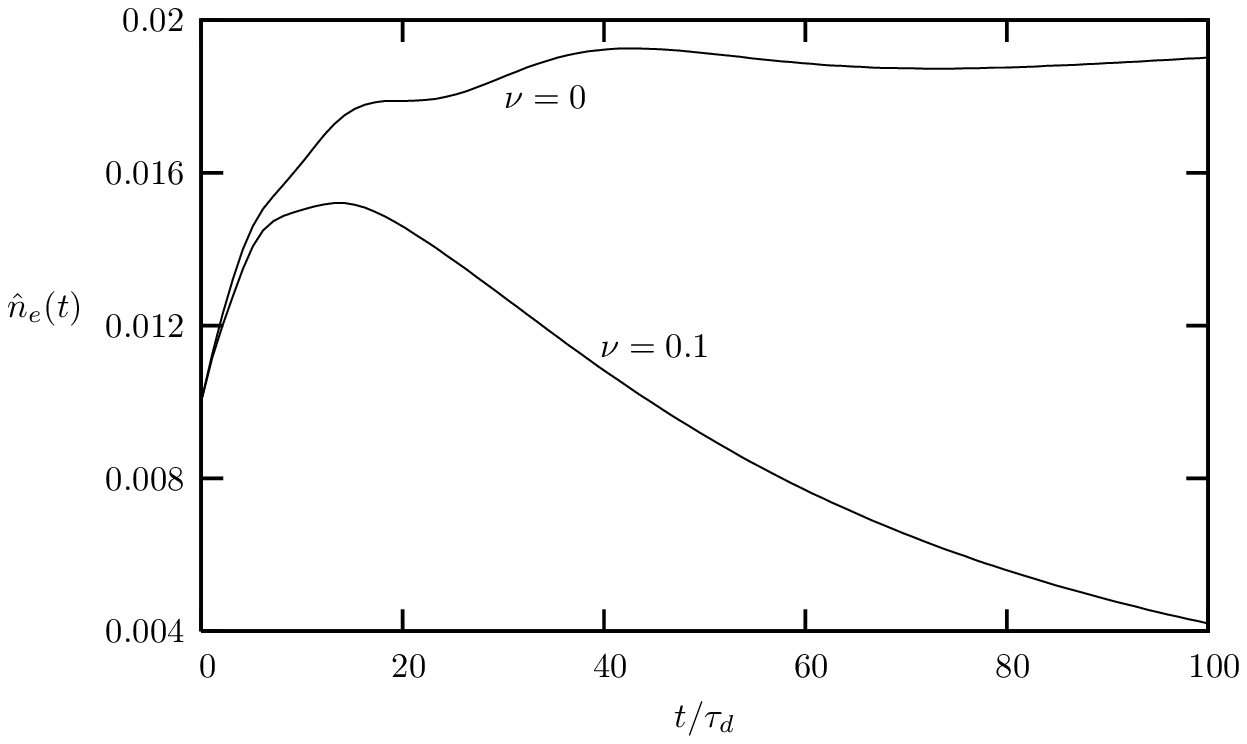}\\
\includegraphics[width=0.48\textwidth]{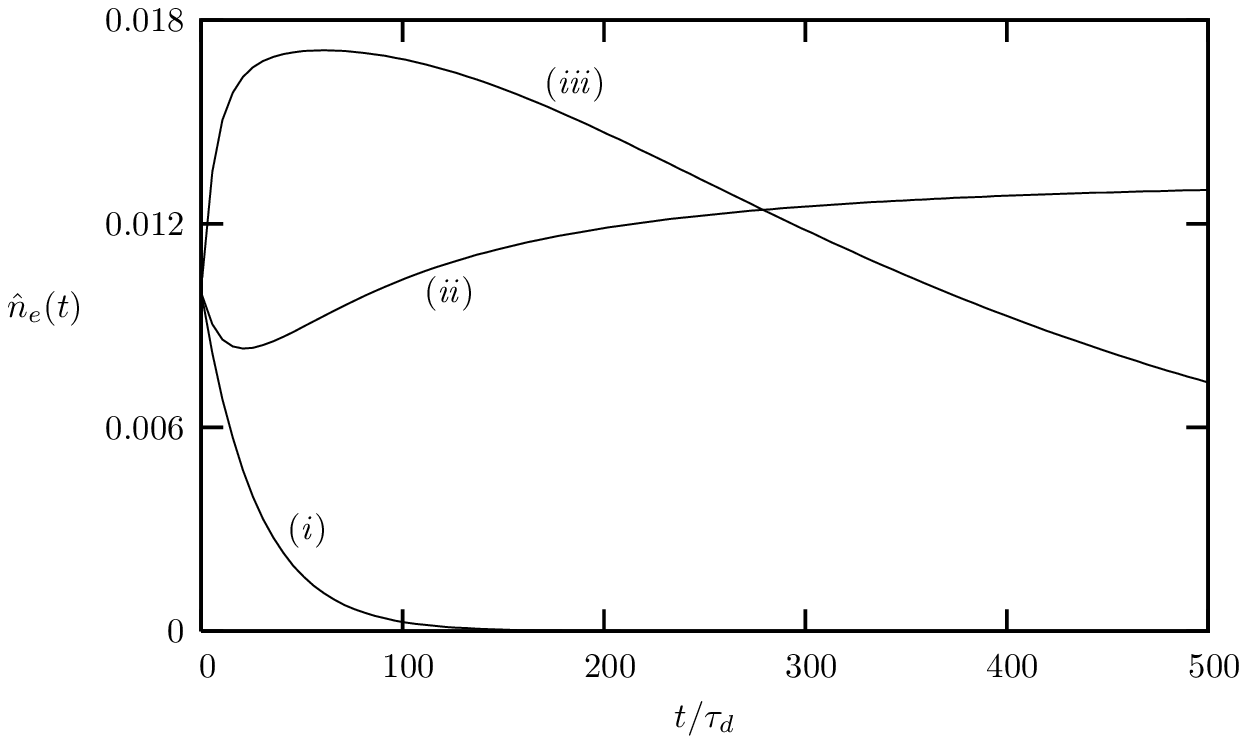}\hfill{}\includegraphics[width=0.48\textwidth]{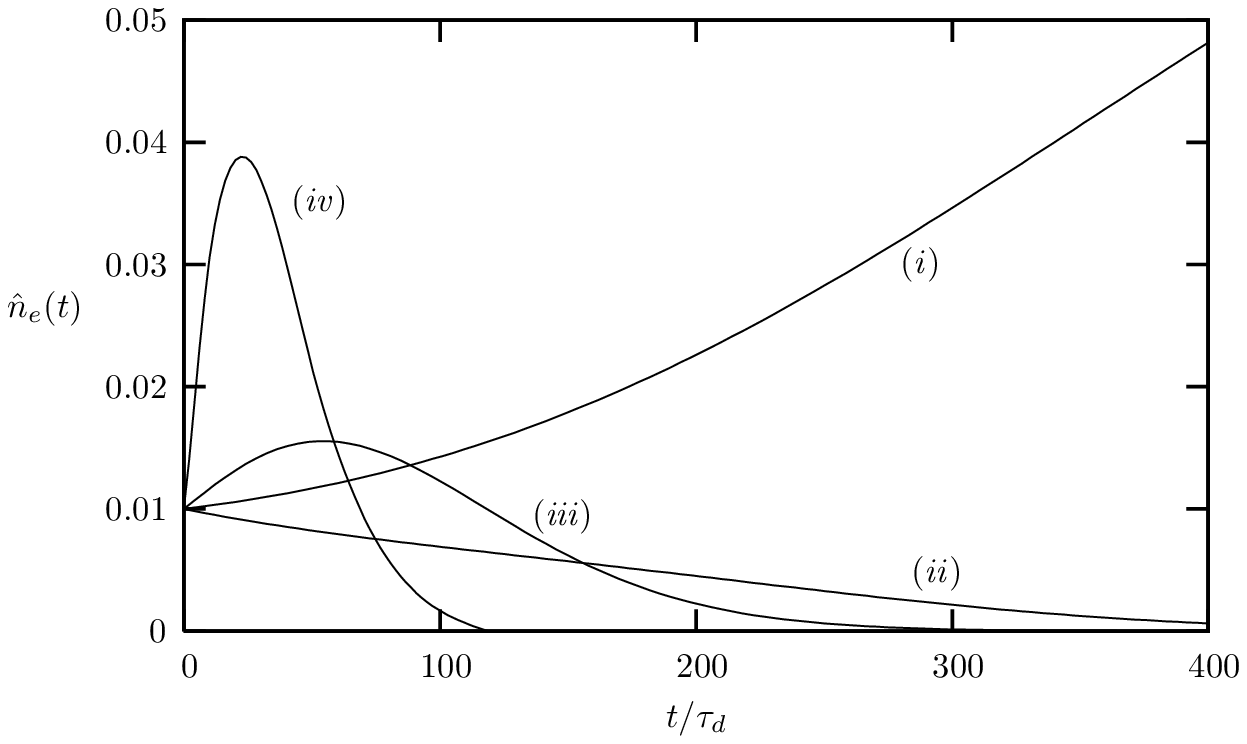}\\

\par\end{centering}

\caption{\label{fig:charge_fluctuation}Row-wise from the top --- (\emph{a})
Low-frequency response of condensation mode for low cooling and expansion,
$\gamma_{c}-\gamma_{h}=10^{-3},\alpha=10^{-3}$ and (\emph{b}) high
cooling and expansion, $\gamma_{c}-\gamma_{h}=0.1,\alpha=0.1$. (\emph{c})
High frequency response to the condensation mode (\emph{i}) $\alpha=0,\gamma_{c}-\gamma_{h}=0,\nu=0.1$,
(\emph{ii}) $\alpha=0.01,\gamma_{c}-\gamma_{h}=0.01,\nu=0.1$, (\emph{iii})
$\alpha=0.1,\gamma_{c}-\gamma_{h}=0.1,\nu=0.1$. (\emph{d}) Low frequency
response to the sound mode (\emph{i}) $\alpha=10^{-3},\gamma_{c}-\gamma_{h}=10^{-3},\nu=0$,
(\emph{ii}) $\alpha=10^{-3},\gamma_{c}-\gamma_{h}=10^{-3},\nu=0.01$,
(\emph{iii}) $\alpha=0.01,\gamma_{c}-\gamma_{h}=0.01,\nu=0.01$, (\emph{iv})
$\alpha=0.1,\gamma_{c}-\gamma_{h}=0.1,\nu=0.01$.}
\end{figure}

In order to facilitate equilibrium balance between radiation loss
and heating, we assume a form of the net cooling function with both
heating and cooling having same dependence on density \cite{fi-1},\begin{equation}
\mathcal{L}_{e0}=f_{1}(n_{e0})[f_{2}(T_{e0})-f_{3}(T_{e0})],\label{eq:heat-loss-function}\end{equation}
and the equilibrium temperature $T_{0}$ is determined by the condition
$\mathcal{L}_{e0}=0$. With the heat-loss function given by Eq.(\ref{eq:heat-loss-function}),
the isobaric stability condition becomes $f_{2}'(T_{e0})>f_{3}'(T_{e0})$.
Without loss of any generality, we assume that $f_{2}\propto T_{e0}$,
$f_{3}\propto T_{e0}^{2}$, and $f_{1}\propto n_{e0}$ so that the
net heat-loss function can be written as,\begin{equation}
\mathcal{L}_{e0}=n_{e0}(C_{1}T_{e0}-C_{2}T_{e0}^{2})=\Lambda-\Gamma,\end{equation}
where $C_{1,2}$ are proportionality constants and $\Lambda$ and
$\Gamma$ are equivalent equilibrium cooling and heating. The cooling
time-scale in terms of $\omega_{c0}$ can be written as\begin{equation}
\omega_{c0}=\omega_{0}\frac{n_{e0}}{n_{0}}\left(\gamma_{c}-\gamma_{h}\frac{T_{e0}}{T_{0}}\right),\end{equation}
with $\gamma_{c,h}$ being the normalised cooling and heating functions
and $n_{0}=n_{e0}|_{t=0}$.

The comprehensive effect of dust-charge fluctuation on the perturbation
amplitude is shown in Fig.\ref{fig:charge_fluctuation}. In the top
panel of Fig.\ref{fig:charge_fluctuation}, the low-frequency response
of the condensation mode is shown for low cooling and slow expansion
(first diagram) and high cooling and rapid expansion (second diagram).
Though the effect of charge-fluctuation is to stabilize the mode,
the effect is more pronounced in the case of rapid expansion as can
be seen from the level of perturbation amplitude. For charge fluctuation
frequency as low as $10\%$ of the dust-acoustic frequency, the condensation
mode is completely stabilized beyond $20\tau_{d}$. As background
expansion tends to enhance the effect of cooling, we have assumed
the cooling and expansion time-scales to be of the same order. The
high-frequency response of the condensation mode is shown in the first
diagram of the lower panel of Fig.\ref{fig:charge_fluctuation} where
the effect of dust-charge fluctuation is shown for different rates
of cooling and expansion. We observe that with finite dust-charge
fluctuation, expansion and equilibrium cooling actually destabilise
the condensation mode {[}curve (\emph{ii}) in Fig.\ref{fig:charge_fluctuation}(c)].
However rapid expansion leads to stabilisation. This is a distinct
departure from the behaviour of the mode in case of low-frequency
perturbation.

The behaviour of the acoustic mode with low-frequency perturbation
is shown in the second figure of the lower panel of Fig.\ref{fig:charge_fluctuation}.
Increasing equilibrium cooling and expansion tend to stabilise the
acoustic mode with finite dust-charge fluctuation. High-frequency
perturbations of the acoustic mode are always damped.

\section{Conclusion}

We have considered the the role of thermal instability of an ionized
plasma with an expanding equilibrium and net equilibrium cooling,
which has a considerable presence of dust and discussed the outcome
of the analysis in the context of expanding planetary nebulae. We
have found that the cumulative effect of equilibrium cooling, background
expansion and the electron capturing by the massive dust particles
is to destabilize the condensation and the acoustic modes which reduces
the cut-off wavelengths beyond which the modes are stable. Therefore,
we see that as a planetary nebula evolves due to expansion, the microstructures
eventually die out, which is in agreement with the observational reports
that these microstructures are transient compared to the evolutionary
life of a nebula. We have shown that the presence of dust particles
in this ionized environment tends to destabilize the acoustic mode.

However, the fluctuation of dust-charge severely damps these instabilities
and the modes are completely stabilized for sufficiently large fluctuation
frequency. As there are no conclusive report on the dust-charge fluctuation
parameter in these ionized plasmas, it will be too much far fetched
to say that the dust-charge fluctuation completely suppresses the
formation of the microstructures, which can be formed out of radiation
induced instabilities. It will be of much interest to view these results
with newer experimental data.

\section*{Appendix}

In this appendix, we show that for non-static equilibrium, it is essential
for the normal mode of the system to be expressed exclusively as a
function of time and a decomposition \cite{nejad} of the kind expressed
in Eq.(\ref{eq:exp_decomp}) is not fruitful. As an example, we take
the case represented in Fig.1, for $k=0.1k_{c}$ for condensation
modes and $k=5k_{c}$ for acoustic modes. We consider, only the effect
of equilibrium expansion in the limit of no net equilibrium cooling.
Thus the equilibrium evolution of the temperature is due to expansion
only {[}see Eq.(\ref{eq:equil_temp})]. In Fig.6, we show the growth
of the perturbed dust density $\hat{n}_{d}(t)$ as obtained by the
full numerical solution of Eqs.(\ref{eq:pert_continuity})--(\ref{eq:pert_ion_density}),
(\ref{eq:pert_temp}), (\ref{eq:quasineutrality}), (\ref{eq:pert_dust_charging}),
and (\ref{eq:pert_dust_momentum}) and the solution through the WKB
expansion, Eq.(\ref{eq:wkb_dr}). The solution, as given by the prescription
represented in Eq.(\ref{eq:exp_decomp}) is plotted in the figures
as dashed lines. As can be seen, the solution from the normal mode
decomposition has a large deviation from the exact numerical solution,
in terms of growth rates, whereas, the WKB solution almost coincides
with the numerical solution. Also note the difference in phases of
the wave mode solution from the numerical one, while the WKB solution
has been able to keep in phase with the numerical solution.%
\begin{figure}[t]
\begin{centering}
\includegraphics[width=0.48\textwidth]{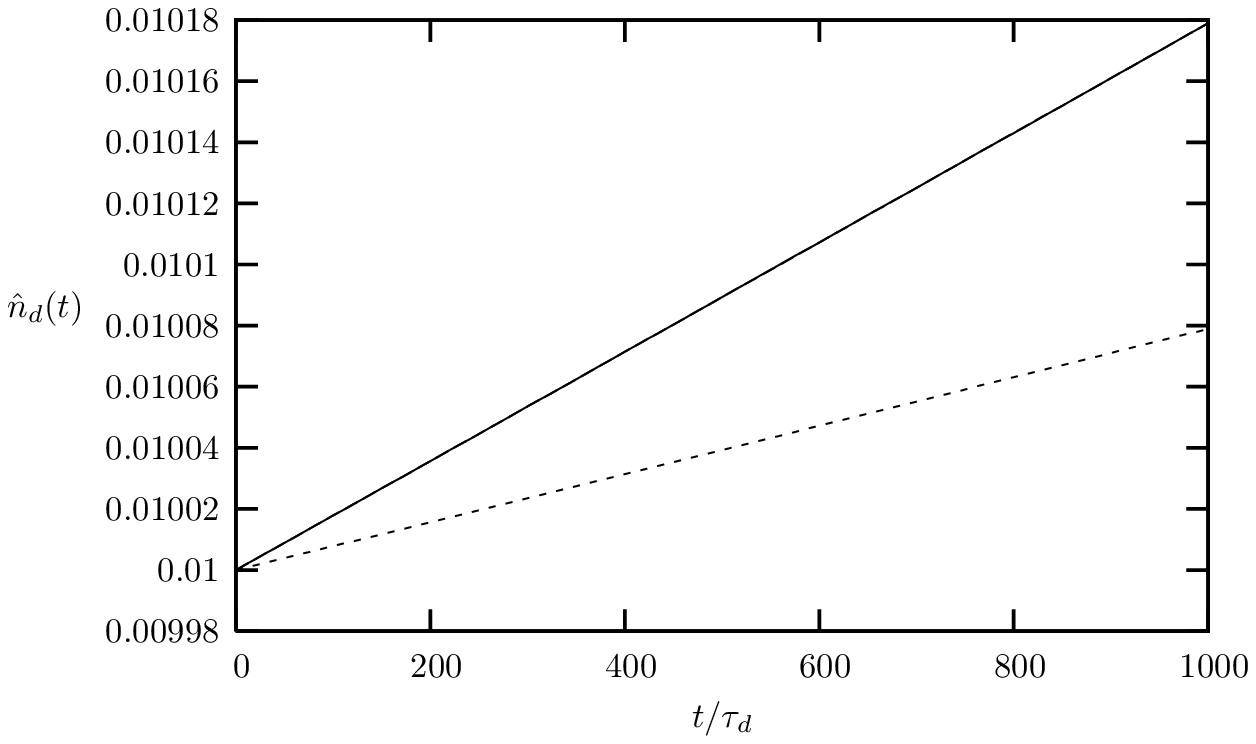}\hfill{}\includegraphics[width=0.48\textwidth]{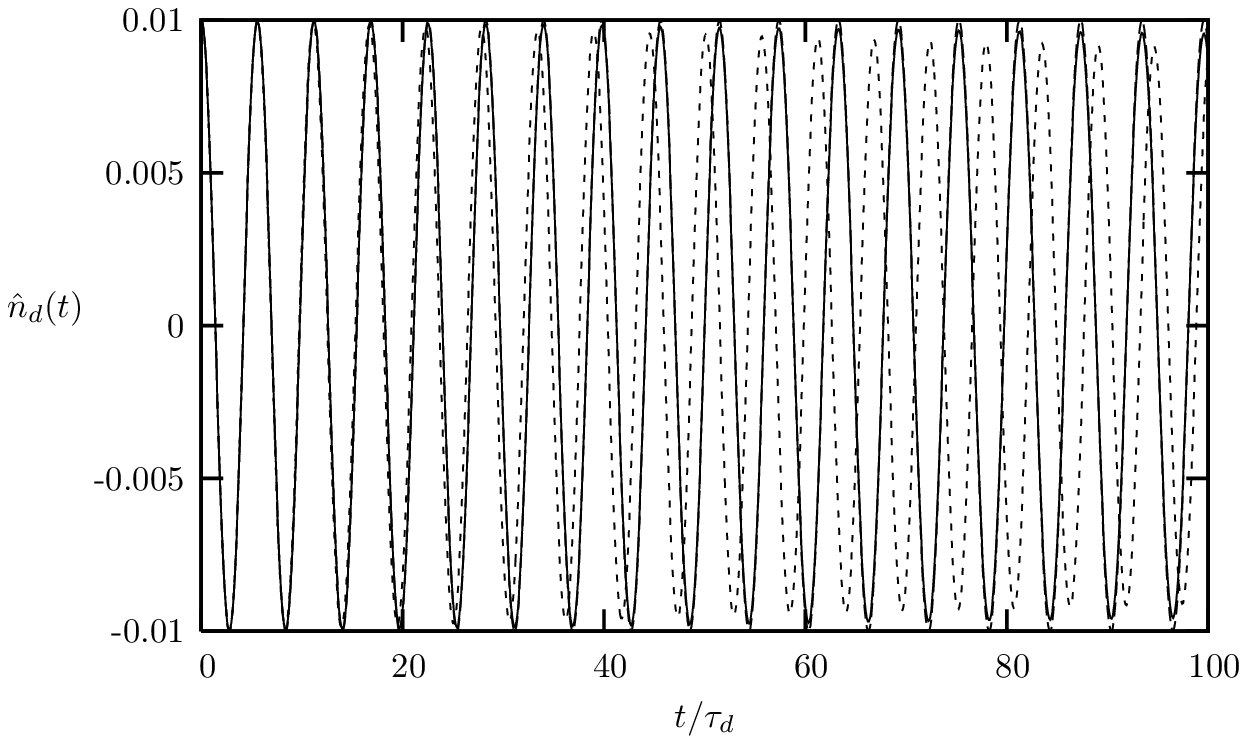}
\par\end{centering}

\caption{The numerical solution of Eqs.(\ref{eq:pert_continuity})--(\ref{eq:pert_ion_density}),
(\ref{eq:pert_temp}), (\ref{eq:quasineutrality}), (\ref{eq:pert_dust_charging}),
and (\ref{eq:pert_dust_momentum}), WKB solution, and the decomposed
solution (shown as dashed lines). The first figure is for condensation
mode and the second one is for wave mode. The solid curves represent
the numerical and WKB solutions which are almost indistinguishable.
In the same parameter regime, the decomposed solution has a large
deviation in amplitudes and phases.}
\end{figure}


\begin{thebibliography}{10}
\bibitem{pa-1}E. N. Parker, Astrophys. J. \textbf{117}, 431 (1953).

\bibitem{we-1}R. Weymann, Astrophys. J. \textbf{132}, 452 (1960).

\bibitem{fi-1}G. B. Field, Astrophys. J. \textbf{142}, 531 (1965).

\bibitem{go-1}A. J. Gomez-Pelaez and F. Moreno-Insertis, Astrophys.
J. \textbf{569}, 766 (2002).

\bibitem{ste-1}C. D. C. Steele, M. H. Ib\'a\~nez, and E. Sira, Phys.
Plasmas \textbf{7}, 3781 (2000).

\bibitem{baek-1}C. H. Baek, H. Kang, J. Kim, and D. Ryu, Astrophys.
J. \textbf{630}, 689 (2005).

\bibitem{baek-2}C. H. Baek, D. Ryu, H. Kang, and J. Kim, Astrophys.
J. \textbf{643}, L83 (2006).

\bibitem{pointek-1}R. A. Piontek and E. C. Ostriker, Astrophys. J.
\textbf{601}, 905 (2004).

\bibitem{pointek-2}R. A. Piontek and E. C. Ostriker, Astrophys. J.
\textbf{629}, 849 (2005).

\bibitem{sh-1}J. C. Shields and R. C. Kennicutt, Astrophys. J. \textbf{454},
807 (1995).

\bibitem{do-1}M. A. Dopita and R. S. Sutherland, Astrophys. J. \textbf{539},
742 (2000).

\bibitem{ol-1}S. Oliveria and W. J. Maciel, Astrophys. Sp. Sc. \textbf{126},
211 (1986).

\bibitem{ba-1}J. A. Baldwin, et. al., Astrophys. J. \textbf{374},
580 (1991).

\bibitem{jana}M. R. Jana, A. Sen, and P. K. Kaw, Phys. Rev. E 48,
3930 (1993).

\bibitem{bora}M. P. Bora, Phys. Plasmas \textbf{11}, 523 (2004).

\bibitem{phillips-1}J. P. Phillips, Astron. Astrophys. \textbf{393},
1027 (2002).

\bibitem{schon-1}D. Schönberner, et. al., Astron. Astrophys. \textbf{431},
963 (2005).

\bibitem{sta-1}G. Stasinska and R. Szczerba, arXiv.astro-ph/9911006
v1 (2006).

\bibitem{pot-1}S. R. Pottasch, \emph{Planetary Nebulae} (D. Reidel
Publ. Comp. 1984).

\bibitem{lenz-1}P. Lenzuni, et. al., Astrophys. J. \textbf{345},
306 (1989).

\bibitem{ki-1}J. Kingdon, G. J. Ferland, and W. A. Fiebelman, Astrophys.
J. \textbf{439}, 793 (1995).

\bibitem{ki-2}J. Kingdon and G. J. Ferland, Astrophys. J. \textbf{477},
732 (1997).

\bibitem{va-1}P. A. M. van Hoof, et. al., Astrophys. J. \textbf{532},
384 (2000).

\bibitem{volk-1}K. Volk, H. Dinerstein, and C. Sneden, in \emph{Planetary
Nebuae}, Proc. IAU Symp. 180, p. 284, 1997.

\bibitem{fi-2}G. B. Field, Astrophys. J. \textbf{187}, 453 (1974).

\bibitem{okro-1}V. A. Okorokov et. al., Astron. Astrophys. \textbf{142},
441 (1985).

\bibitem{mar-1}H. Marten, R. Szczerba, and Th. Blöcker, in \emph{Planetary
Nebuae} (Proc. IAU Symp. 155, 1993), p. 363.

\bibitem{pandey}B. P. Pandey, V. Krishan, and M. Roy, Pramana \textbf{56},
95 (2001).

\bibitem{planetary}B. Balick et. al., Astrophys. J. \textbf{424},
800 (1994).

\bibitem{wein}S. Weinberg, \emph{Gravitation and Cosmology: Principles
and Applications of the General Theory of Relativity} (Wiley, 2001).

\bibitem{bender}C. M. Bender and S. A. Orszag, \emph{Advanced Mathematical
Methods for Scientists and Engineers} (McGraw Hill, New York, 1978).

\bibitem{nejad}M. Nejad-Asghar and J. Ghanbari, arXiv:astro-ph/0603172
v2 (2006).

\bibitem{spitzer}L. Spitzer, \emph{Physics of Fully Ionized Gases}
(Wiley, New York, 1962).

\bibitem{rybicki}G. B. Rybicki and A. P. Lightman, \emph{Radiative
Precesses in Astrophysics} (Wiley, New York, 1979).

\bibitem{mendis}D. A. Mendis and M. Rosenberg, Ann. Rev. Astron.
Astrophys. \textbf{32}, 419 (1994).

\bibitem{guer-1}M. A. Guerrero, E. Villaver, and A. Manchado, Astrophys.
J. \textbf{507}, 889 (1998).

\bibitem{stran-1}L. Stanghellini and J. B. Kaler, Astrophys. J. \textbf{343},
811 (1995).

\bibitem{chou-1}Y.-H. Chou, in \emph{Planetary Nebulae}, edited by
S. Torres-Peimbert (IAU Symp. 131, Kluwer, Dordrecht, 1989), p. 105.
\end{thebibliography}
\end{document}